\documentclass[conference,a4paper]{IEEEtran}
\usepackage[utf8]{inputenc}
\usepackage{amsmath}
\usepackage{amsfonts}
\usepackage{amssymb}
\usepackage{graphicx}
\usepackage{xcolor}
\usepackage{soul}
\usepackage{comment}
\usepackage{caption}
\usepackage{subcaption}
\usepackage{balance}

\usepackage{epsfig}    
\usepackage{multirow}
\usepackage{psfrag}
\usepackage[linesnumbered,ruled]{algorithm2e}
\usepackage{float}
\usepackage{diagbox}
\usepackage{url}
\usepackage{hyperref}
\usepackage{xcolor}
\usepackage{stfloats}
\usepackage{cite}
\usepackage[normalem]{ulem}

\usepackage{tabulary}

\definecolor{boristext}{rgb}{0.22, 0.44, 0.88}
\definecolor{boriscomments}{rgb}{0.88, 0.04, 0.04}
\definecolor{boristochange}{rgb}{0.2, 0.8, 0.8}

\title{Understanding Multi-link Operation in Wi-Fi 7:\\Performance, Anomalies, and Solutions}

\author{
\IEEEauthorblockN{Marc Carrascosa-Zamacois, Giovanni Geraci, Lorenzo Galati-Giordano, Anders Jonsson,
and Boris Bellalta}
\thanks{M. Carrascosa-Zamacois, G. Geraci, A. Jonsson, and B. Bellalta are with Univ. Pompeu Fabra, Barcelona, Spain. They were supported in part by grants PID2021-123995NB-I00, PGC2018-099959-B-I00, PRE2019-088690, RTI2018-101040-A-I00, PID2021-123999OB-I00, and by the ``Ram\'{o}n y Cajal" program. L. Galati-Giordano is with Nokia Bell Labs, Stuttgart, Germany.}
}

\IEEEoverridecommandlockouts
\begin{document}

\maketitle

\begin{abstract}
Will Wi-Fi 7, conceived to support extremely high throughput, also deliver consistently low delay? 
The best hope seems to lie in allowing next-generation devices to access multiple channels via multi-link operation (MLO). 
In this paper, we aim to advance the understanding of MLO, placing the spotlight on its packet delay performance. 
We show that MLO devices can take advantage of multiple contention-free links to significantly reduce their transmission time, but also that they can occasionally starve one another and surprisingly incur a higher delay than that of a well planned legacy single link operation. 
We next examine and explain this anomaly, also putting forth practical workarounds to circumvent it. 
We conclude by pointing to other disruptive features that, if successfully paired with MLO, can usher in exciting and unprecedented opportunities for Wi-Fi 8.
\end{abstract}

\section{Introduction}

As we head out of a pandemic that made connectivity truly dependable, our appetite for mobile data is stronger than ever. Myriad engineers behind the development of Wi-Fi, the technology carrying nearly two thirds of all wireless data, relentlessly feed this hunger by crafting ever more clever amendments, defining new Wi-Fi generations one after another. At the time of writing, Wi-Fi 6 and 6E are a commercial reality, the making of Wi-Fi 7 is nearing completion, and the definition of Wi-Fi 8 starts catalyzing the interest of tech giants and avid researchers alike \cite{lopez2019ieee,garcia2021ieee,ResCor22}. Yet before debating or fantasizing about what Wi-Fi 8 should be, what will Wi-Fi~7 deliver?

The IEEE 802.11be amendment, expected to be at the heart of Wi-Fi 7, will remain loyal to its legacy---and to its very name: EHT, short for `Extremely High Throughput'---by augmenting data rates through various upgrades ranging from wider bandwidths (up to 320 MHz)

to higher modulation orders (up to 4096-QAM) \cite{khorov2020current,deng2020ieee,yang2020survey}. But besides features boosting the nominal throughput, many experts point to multi-link operation (MLO) as the true paradigm shift Wi-Fi 7 will bring to the table. MLO will allow Wi-Fi devices to concurrently operate on multiple channels through a single connection, aiming to support applications demanding not only higher capacity but also lower delay~\cite{CheCheDas22,lopez2022multi}.

Unlike merely multiplying the peak throughput gains provided by scaling up bandwidth and spectral efficiency, quantifying the advantages brought about by MLO in realistic scenarios is no straightforward endeavor. And while several works have recently made valuable attempts at studying how MLO performs in terms of  throughput and delay \cite{naik2021can,murti2022multilink,lacalle2021analysis,naribole2020simultaneous,suer2022adaptive,song2020performance}, a deep and widespread understanding of the latter remains little more than wishful thinking. Indeed, the exact benefits on a device employing MLO for delay-sensitive applications and the effects on coexisting basic service sets (BSSs) hinge on the specific MLO implementation, with several being defined in 802.11be to trade off complexity and flexibility. Furthermore, as we will show in later sections, these benefits---or the lack thereof---highly depend on the traffic load, the surrounding environment, and the channel allocation strategy adopted.

In this paper, we shed light on the delay performance of STR EMLMR (standing for `Simultaneous Transmit and Receive Enhanced Multi-link Multi-radio'), arguably the most flexible MLO mode, under varying traffic demand, congestion, and channel allocation strategies. We explain and quantify its main virtues with respect to legacy single-link (SL) as well as its caveats, also putting forth possible solutions to the latter. Our main takeaways can be summarized as follows:
\begin{itemize}
\item 
In scenarios devoid of contention, STR EMLMR exploits additional available links to perform multiple transmissions in parallel, proportionally reducing the channel access delay.

\item 
In the presence of high load and contention, STR EMLMR devices frequently access multiple links thereby blocking contending neighbors, occasionally causing larger delays than those experienced with a static SL channel assignment.
\item 
For consistent worst-case delay reduction, STR EMLMR may require more channels than contending BSSs and/or performing a clever channel assignment that entirely circumvents delay anomalies caused by sporadic BSS starvation. 

\end{itemize}

Compared to existing work, the novelty and contribution of the present paper is at least threefold:
\begin{itemize}
    \item We introduce the reader to MLO, review its different implementations, and illustrate, for the first time, the intricate interactions it triggers between contending BSSs. We demonstrate how such interplay may turn out being benign or unfavorable, depending on the traffic load and channel allocation strategies.
    \item We identify, quantify, and explain, through novel results, the delay anomalies that may surprisingly arise when employing STR EMLMR in the presence of high load and contention. We also propose multiple solutions to circumvent such anomalies and we evaluate and compare their effectiveness.
    \item We provide a concise yet complete picture of the virtues and caveats of MLO by putting our findings into an even broader context. We also invoke follow-up research on other disruptive features that, if successfully paired with MLO, can usher in exciting and unprecedented opportunities for Wi-Fi 8.
\end{itemize}
\begin{figure*}
    \centering
    \includegraphics[width=\textwidth]{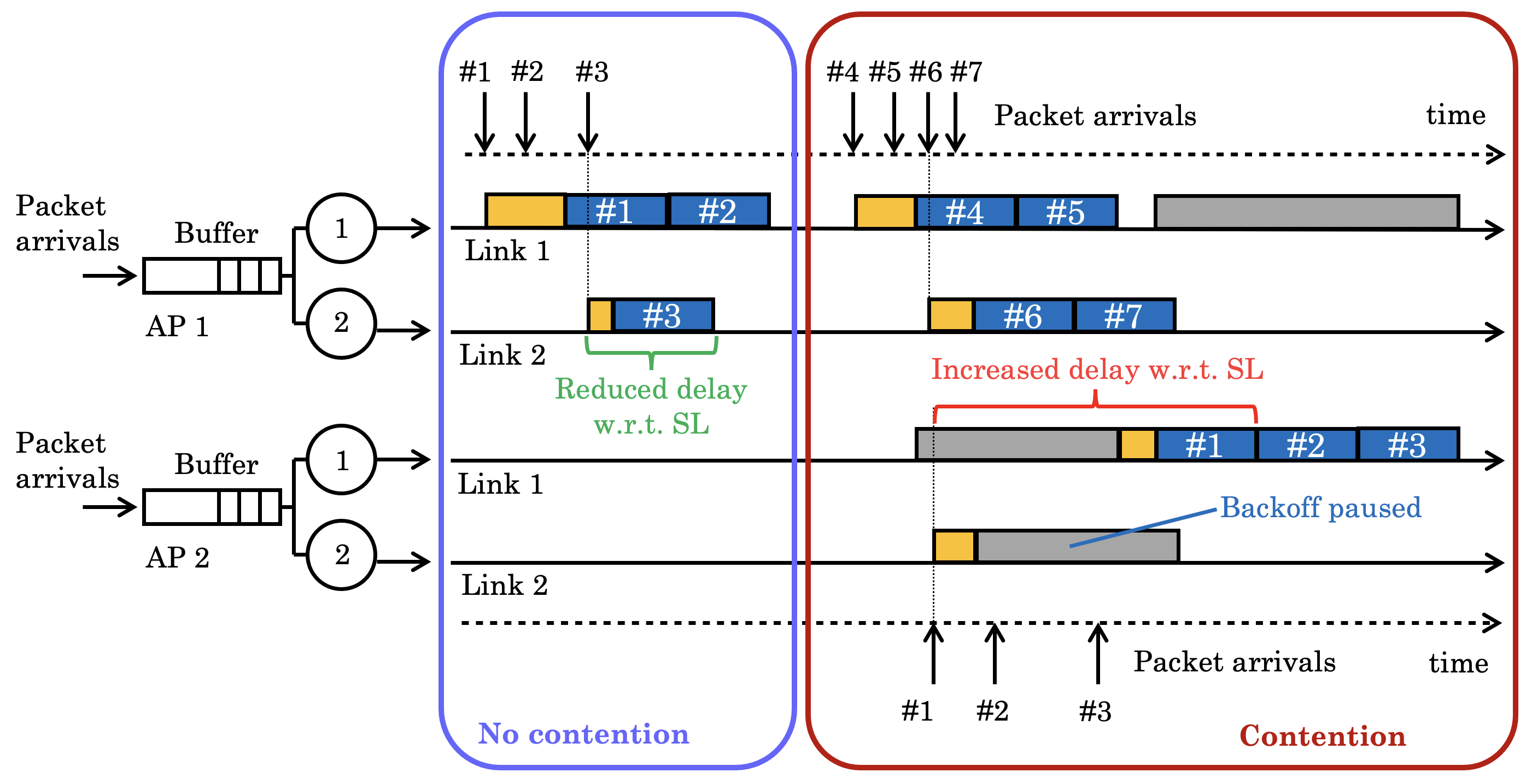}
    \caption{Example of STR EMLMR operations and packet interactions without (left) and with (right) contention. Grey, orange, and blue slots denote occupied channels, ongoing backoffs, and successful transmissions, respectively. Consecutive blue slots indicate aggregated packets. For illustration purposes, all transmissions are downlink and the corresponding ACKs are omitted. In the example, for AP 1, packet \#3 experiences a lower delay than it would under SL operations. For AP 2 instead, packet \#1 undergoes a higher delay than it would with SL.
    }
    \label{fig:diagram}
\end{figure*}

\section{A Primer on Multi-link Operation}

In addition to legacy \textsl{SL} (Single-link) channel access---which follows the 802.11 Distributed Coordination Function~(DCF) over one channel---Wi-Fi 7 will allow MLO through single association, with channel contention and access performed independently for each link.

The 802.11be amendment defines different MLO implementation flavors, with the main ones summarized as follows \cite{CheCheDas22}.

\subsection{Multi-link Flavors}

\subsubsection*{Enhanced Multi-link Single-radio (EMLSR)} EMLSR enables a single-radio multi-link device (MLD) to listen to two or more links simultaneously, e.g., by splitting its multiple antennas, performing clear channel assessment and receiving a limited type of control frames. EMLSR supports opportunistic spectrum access at a reduced cost, as it requires a single fully functional 802.11be radio plus several other low-capability radios able only to decode 802.11 control frame preambles. Upon reception of an initial control frame on one of the links, the EMLSR MLD can switch to the latter and operate using all antennas.

\subsubsection*{Enhanced Multi-link Multi-radio (EMLMR)}
For a MLD implementing EMLMR, all radios are 802.11be-compliant and allow operating on multiple links concurrently. 
EMLMR is further classified into two modes:
\begin{itemize}
    \item
    \emph{Non-simultaneous Transmit and Receive (NSTR) EMLMR}, where no simultaneous transmission and reception is allowed over a pair of links in order to prevent self interference at the MLD. The latter entails ensuring near alignment in the end time of physical layer protocol data unit that are simultaneously transmitted, so as to avoid that subsequent incoming responses on one link, e.g., ACKs, overlap with the remaining transmission on another link.
    \item
    \emph{Simultaneous Transmit and Receive (STR) EMLMR}, where the above rule does not apply. In order to avoid uplink-to-downlink intra-device interference, operating STR EMLMR requires sufficient frequency separation between the channels used by different links and/or sophisticated self-interference cancellation capabilities. For instance, STR EMLMR with four links, each on an 80~MHz channel, could be implemented by using two channels each in the 5~GHz and 6~GHz bands, with a minimum channel separation of 160~MHz, and equipping MLDs with suitable radio-frequency filters.

\end{itemize}

A remark is in order about the `E' in EMLSR and EMLMR, standing for `enhanced'. Indeed, non-enhanced versions of both have also been defined \cite{CheCheDas22}, their main features summarized as follows:
\begin{itemize}
    \item \textsl{MLSR}, where unlike EMLSR, clear channel assessment and control frame reception (and of course,  data transmission/reception) can only be performed on one channel at a time, thereby limiting opportunistic link selection.
    \item \textsl{MLMR}, which compared to EMLMR only lacks extra capabilities to dynamically reconfigure spatial multiplexing over multiple links. This difference is immaterial for the case studies of the present paper.
\end{itemize}

In the remainder of this article, we place the spotlight on STR EMLMR since it is the MLO operation mode that grants the highest degree of flexibility and requires the least amount of signaling, thus being the most likely to be adopted in first-wave Wi-Fi 7 commercial products. Unlike previous work devoted to the achievable throughput of STR EMLMR, we focus on its delay performance as we deem it crucial to support ever more proliferating real-time applications.

\subsection{A Close-Up of STR EMLMR}

As shown in Fig.~\ref{fig:diagram}, exemplifying STR EMLMR in action over two links, it turns out that this mode of operation can affect the packet delay in multiple ways, depending on the particular scenario at hand. In the following, we provide two examples that illustrate how STR EMLMR can respectively reduce and increase the delay with respect to legacy SL operations. For the latter (not shown), we assume an orthogonal channel assignment as a benchmark, with AP 1 and AP 2 operating on link 1 and link 2 only, respectively.

\subsubsection*{Delay reduction through STR EMLMR}
Let us begin by focusing on the left hand side of Fig.~\ref{fig:diagram}, where AP 2 is inactive and AP 1 can take advantage of two available links by routing traffic to either as needed. In the example, packets \#1 and \#2 are aggregated and promptly transmitted over link 1. As for packet \#3, which arrives during an ongoing transmission, a new backoff is started on link 2, followed by a transmission. Packet \#3 thus enjoys a significant delay reduction compared to a legacy SL scenario, as in the latter it would have needed to wait for the ongoing transmissions on link 1 to be completed. 

\subsubsection*{Delay anomaly in STR EMLMR}
The right hand side of Fig.~\ref{fig:diagram} illustrates a scenario where AP 1 and AP 2, both implementing STR EMLMR, contend for channel access. In this example, AP 1 aggregates packets \#4 and \#5 upon backoff expiration and transmits them over link 1. Meanwhile, more traffic arrives, namely packets \#6 and \#7 at AP 1 and packets \#1 and \#2 and AP 2. Since link 1 is occupied by AP 1, both AP~1 and AP~2 undergo contention for link 2, with the backoff for AP 1 expiring first. AP 1 thus aggregates and transmits packets \#6 and \#7 on link 2, thereby occupying both links concurrently. It is only after the transmission of packets \#4 and \#5 by AP 1 is completed that AP 2 can eventually aggregate and transmit all its queued packets on link 1. In the example, these packets experience a much higher delay than they would have under legacy SL operations. 
Indeed, with SL and a static channel allocation (e.g., AP 1 on link 1 and AP 2 on link 2), AP 1 would have not been able to occupy both links simultaneously, and therefore would have not temporarily forced AP 2 into starvation. We identify this phenomenon as an \emph{anomaly} of MLO, and will devote Section~IV to its better understanding.

As it can be seen through the above two examples, STR EMLMR is capable of taking advantage of multiple links to reduce the channel access time with respect to SL, but also to occasionally starve neighboring BSSs thereby increasing their delay. In the sequel, we will confirm and quantify these two phenomena through targeted simulation campaigns.
\section{STR EMLMR in Contention-free Scenarios}

\begin{figure}
    \includegraphics[width = 0.99\columnwidth]{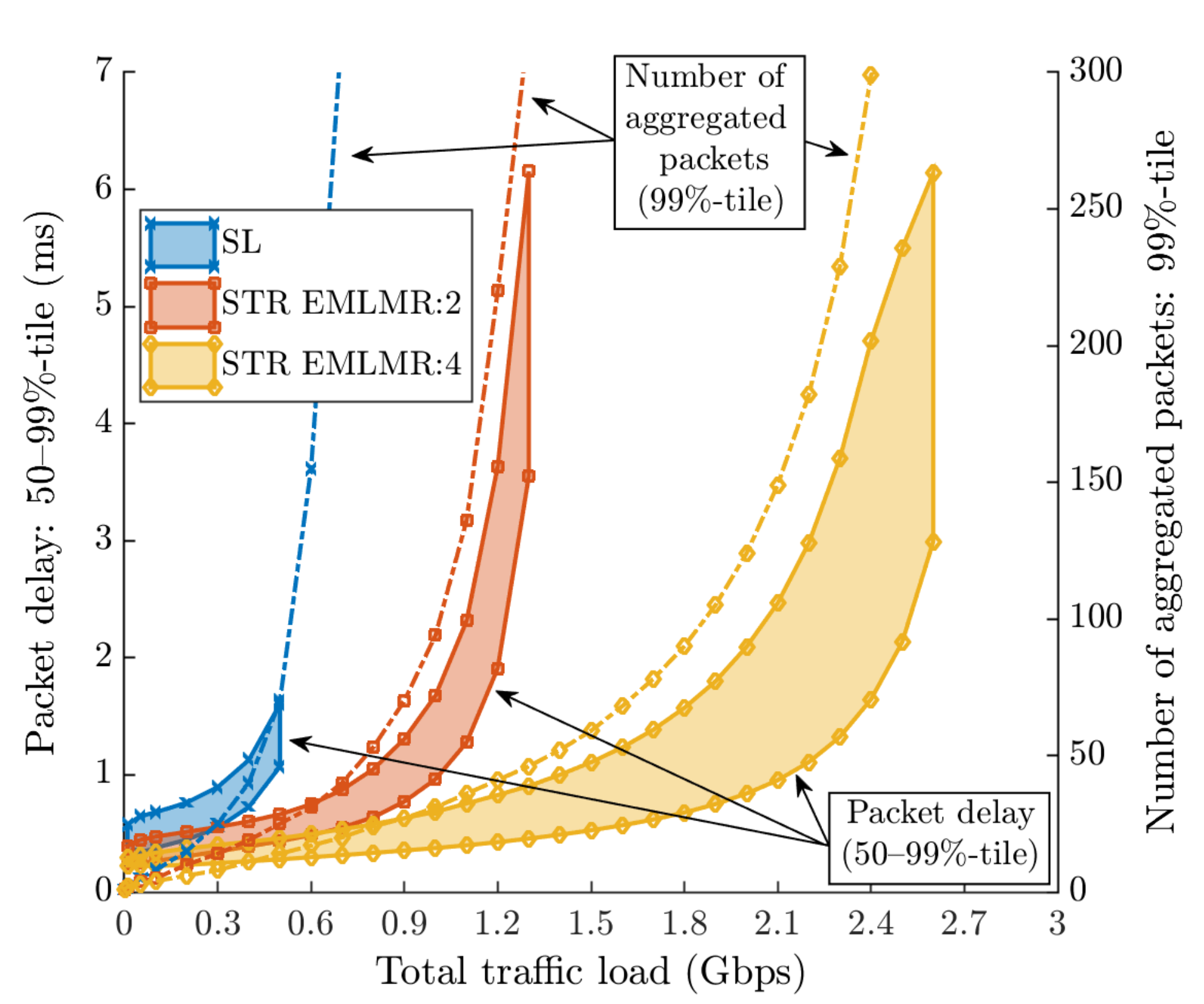}
    \caption{Contention-free scenario: packet delay (spanning 50--99\%-tile) and number of aggregated packets (99\%-tile) vs. traffic load for SL, STR EMLMR:2, and STR EMLMR:4.} 
    \label{Fig:NoContention}
\end{figure}

We begin by considering a single, isolated BSS with one MLD station (STA) associated to an MLD access point (AP), and evaluate the delay performance of STR EMLMR in such a contention-free scenario. Without loss of generality, we focus on downlink traffic and assume Poisson arrivals with constant packet size of 12000 bytes, 80~MHz channels, two spatial streams, and a modulation and coding scheme of 256-QAM 3/4 \cite{CheCheDas22}. Packet aggregation is employed with the number of aggregated packets decided at the start of a transmission, up to a maximum of 1024. A buffer size of 4096 packets is employed, ensuring sufficient room for the maximum allowed number of aggregated packets. For this scenario, we study the effect of the traffic load on the packet delay under three schemes, namely:
(i) \textsl{SL}, taken as the baseline, (ii) \textsl{STR EMLMR:2}, where the isolated AP can use two links at any time, and (iii) \textsl{STR EMLMR:4}, with four links simultaneously available.

\subsubsection*{Packet delay} Fig.~\ref{Fig:NoContention} shows the delay statistics with shaded curves ranging from 50\%-tile to 99\%-tile (i.e., median to 1\%-worst) using SL and STR EMLMR with two or four links.
Intuitively, as more links are available and can be accessed dynamically, a certain delay requirement can be met for proportionally higher values of the traffic load, i.e., while supporting a proportionally higher throughput. For instance, given a median delay of 1~ms, SL, STR EMLMR:2, and STR EMLMR:4 can roughly support up to half, one, and two Gbps, respectively. Similarly, given a certain traffic load, availing of extra links decreases the delay, albeit with diminishing returns. For instance at 0.5~Gbps, the three schemes incur 99\%-tile delays of about 1.6, 0.7, and 0.5~ms. Nonetheless, depending on the traffic load, accessing multiple links may be the only way to prevent the delay from growing unbounded. E.g., a load of 1~Gbps exceeds the capacity of a single channel, thus SL incurs unbounded delay, whereas STR EMLMR:2 and STR EMLMR:4 keep the delay below 1.7 and 0.7~ms, respectively, 99\% of the time.

\subsubsection*{Packet aggregation}
Fig.~\ref{Fig:NoContention} also displays the 99\%-tile for the corresponding number of packets aggregated under each of the three schemes (dashed lines). Even at moderate loads, owing to its inability of using multiple links, SL experiences a higher buffer congestion and is forced to aggregate a much larger number of packets per transmission than STR EMLMR. 
The latter can instead parallelize access on multiple links, reducing the buffer congestion and thus the number of aggregated packets for each transmission.

\subsubsection*{Takeaway} In scenarios devoid of contention, STR EMLMR can exploits extra links---even across different frequency bands, something SL is not capable of---to operate on a wider bandwidth, and can therefore meet a certain delay requirement while supporting higher traffic loads (i.e., throughput) than SL. 

While the above results are somewhat expected, they are in stark contrast to the delay anomaly experienced by MLDs in crowded scenarios, quantified in the next section.

\begin{figure*}
    \centering
    \begin{subfigure}[b]{0.31\textwidth}
    \includegraphics[width = \textwidth]{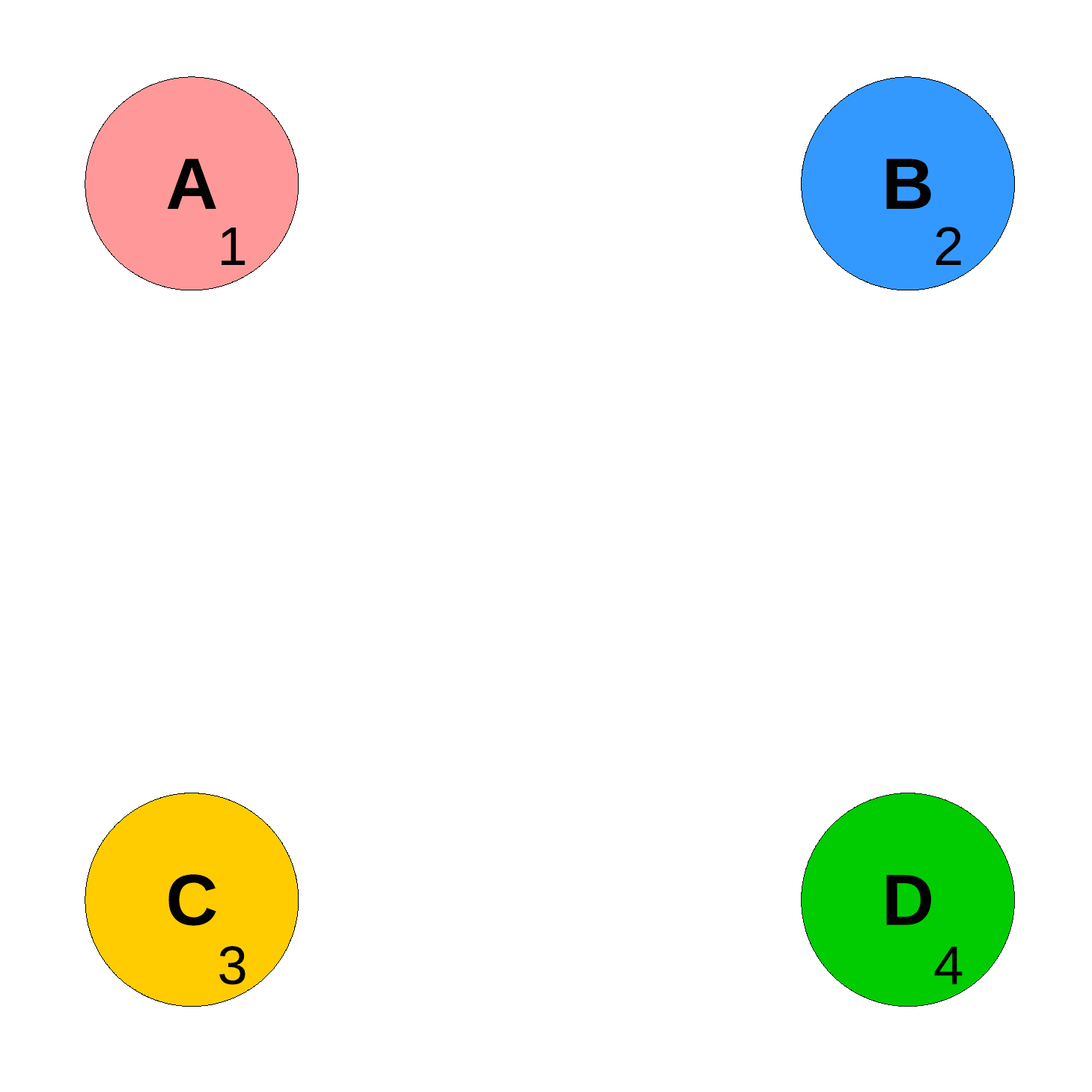}
    \caption{SL, a single 80~MHz channel exclusively assigned to each BSS and no contention.}
    \label{fs1}
    \end{subfigure}
    \hspace{0.02\textwidth}
     \begin{subfigure}[b]{0.31\textwidth}
    \includegraphics[width = \textwidth]{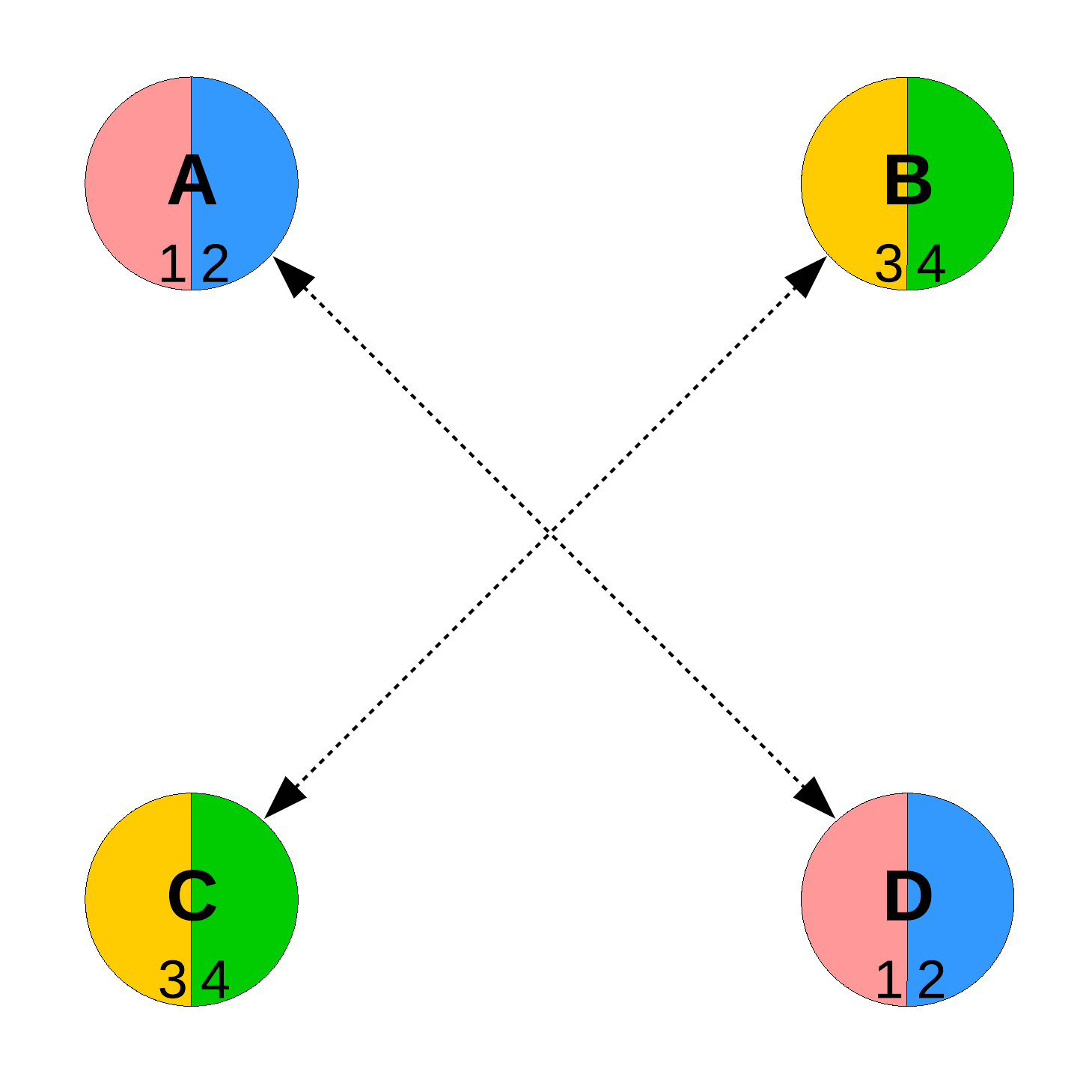}
    \caption{STR EMLMR:2, two 80~MHz channels per BSS, both shared with one more BSS.}
    \label{fs2}
    \end{subfigure}
    \hspace{0.02\textwidth}
     \begin{subfigure}[b]{0.31\textwidth}
    \includegraphics[width = \textwidth]{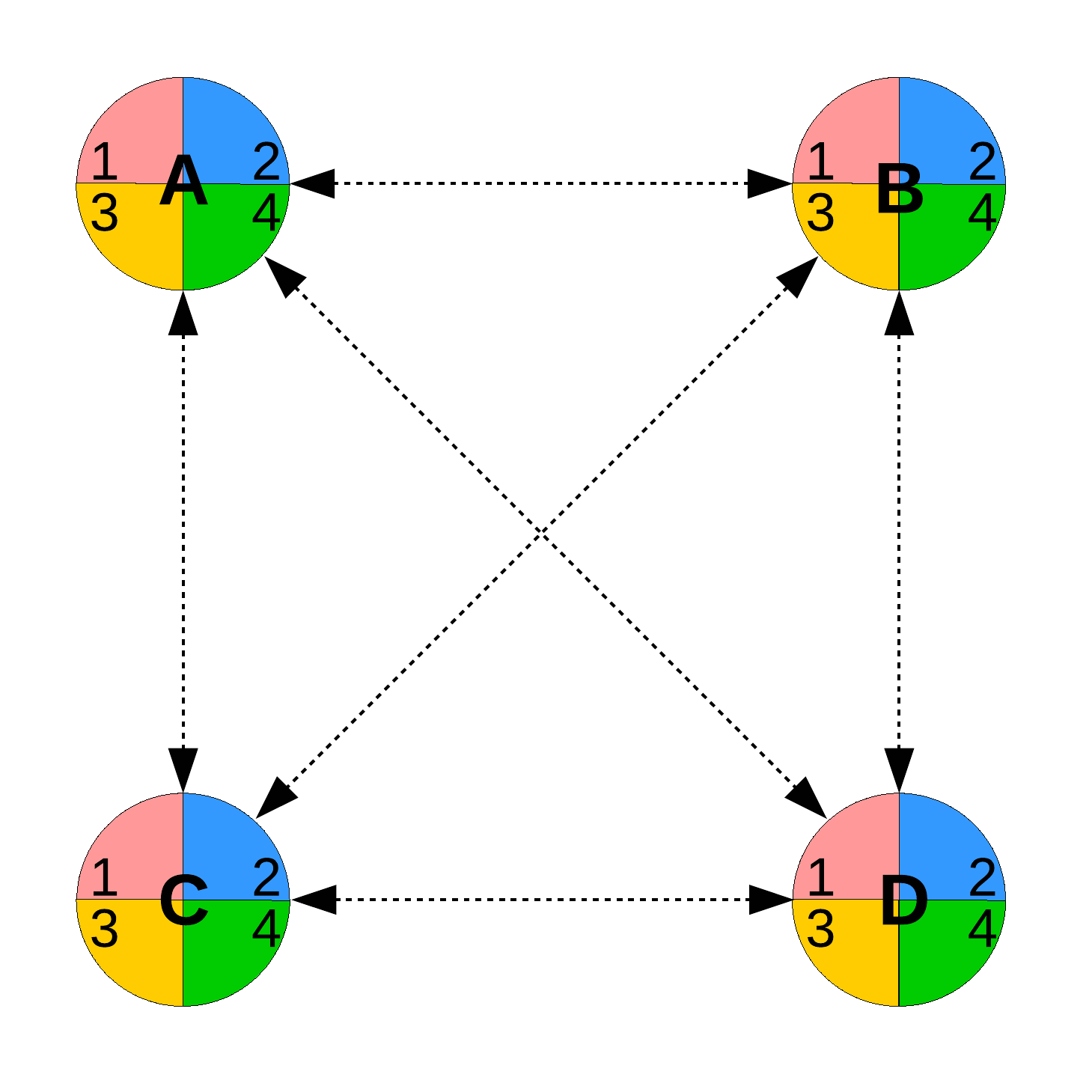}
    \caption{STR EMLMR:4, four 80~MHz channels per BSS, all shared with three more BSSs.}
    \label{fs3}
    \end{subfigure}
    \caption{Three modes of operation considered for a crowded scenario: (a) SL, (b) STR EMLMR:2, and (c) STR EMLMR:4. Colors and numbers refer to different channels, letters denote BSSs, and dashed arrows indicate contention between BSSs.}
    \label{fs}
\end{figure*}
\section{STR EMLMR in Crowded Scenarios}

We now investigate when the delay reduction provided by STR EMLMR is maintained in the presence of contention, and when instead the delay is increased due to the starvation phenomenon, i.e., the \emph{anomaly}, outlined in Section~II-B. To this end, we turn our attention to a more crowded enterprise scenario with 4 BSSs as depicted in Fig.~\ref{fs}. Each BSS comprises one AP and one associated STA, all BSSs are in the coverage range of each other, and the whole system has a limited amount of resources, namely four orthogonal 80~MHz channels. For this challenging scenario, we consider three possible modes of operation, each making a different use of the four available channels:
\begin{itemize}
\item \textsl{SL}, with a single channel exclusively assigned to each BSS, as illustrated in Fig.~\ref{fs1}, and no contention. Again, we take this mode as the baseline to assess STR EMLMR.
\item \textsl{STR EMLMR:2}, as shown in Fig.~\ref{fs2}, where each BSS employs two channels and shares both with one more contending BSS.
\item \textsl{STR EMLMR:4}, as shown in Fig.~\ref{fs3}, where all four BSSs employ and contend for all four channels. 
\end{itemize}
Note that the above three arrangements assume statically assigning channels to BSSs according to a specific reuse scheme, and thus embody a hypothetical enterprise use case.
In this section, we assume the same values of total traffic load as in Section~III, but this time evenly spread among all BSSs, i.e., one quarter each. The scenarios in Section~III (Fig.~\ref{Fig:NoContention}) vs. Section~IV (Fig.~\ref{Fig:Contention}) can thus be regarded as an asymmetric vs. symmetric distribution of the same total load between contending BSSs.

\subsubsection*{Packet delay}
Fig.~\ref{Fig:Contention_Delay} shows the mean, 95\%-tile, and 99\%-tile delay using SL, STR EMLMR:2, and STR EMLMR:4 vs. the total traffic load. 
For a relatively low load of 0.1~Gbps, the delay is decreased by adding multiple links since there is negligible contention and STR EMLMR can quickly find and exploit extra transmission opportunities, as previously shown in Section~III.

However, once the load reaches higher values such as 1~Gbps and above, STR EMLMR worsens the delay compared to SL, and four links incur a higher delay than two. These results stem from the anomaly illustrated on the right hand side of Fig.~\ref{fig:diagram}, and can be further explained by the interplay between multi-link contention and packet aggregation, detailed as follows.

\subsubsection*{Multi-link contention} 

\begin{figure}
    \centering
    \begin{subfigure}[b]{0.99\columnwidth}
    \includegraphics[width=0.99\columnwidth]{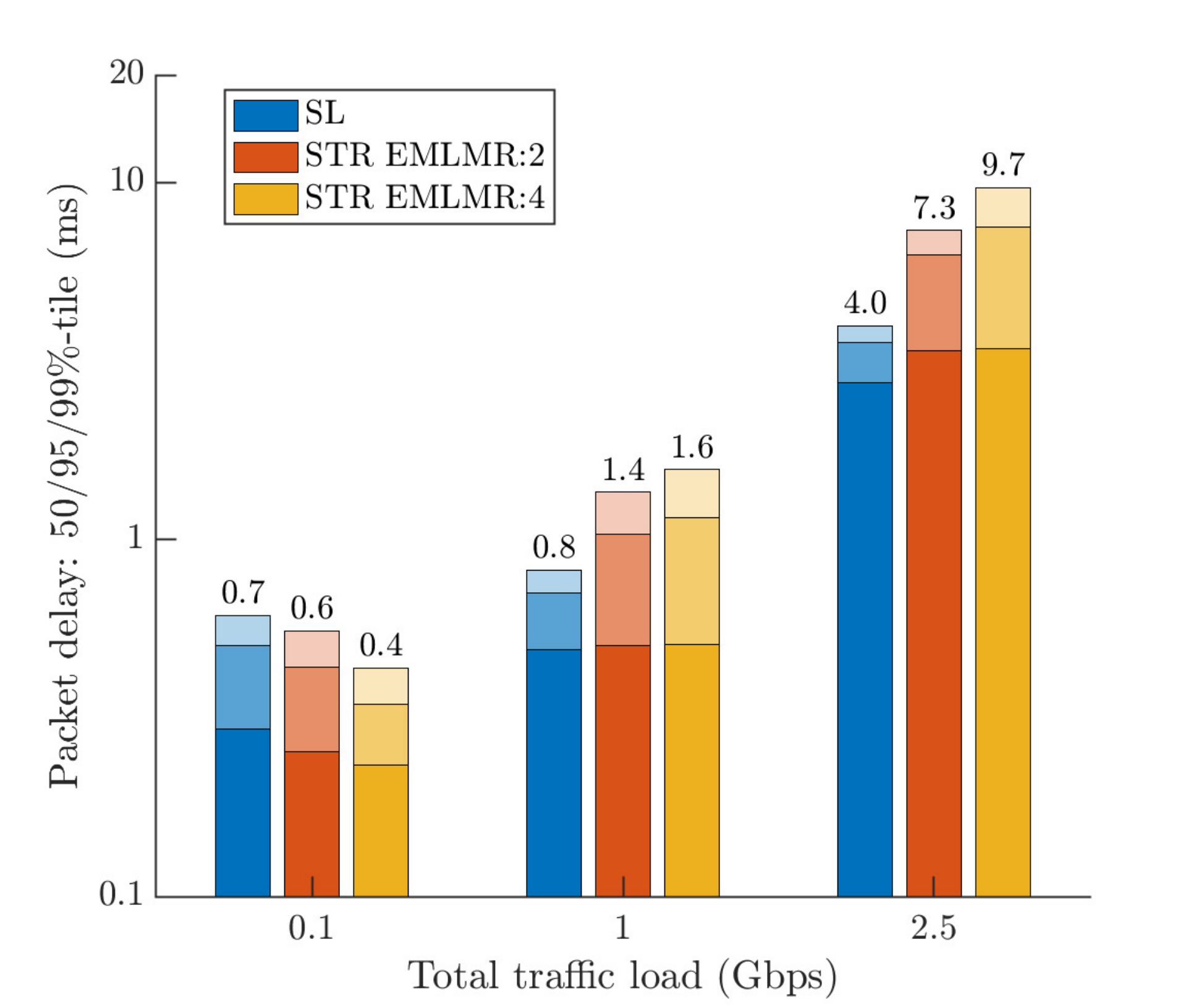}
    \caption{Delay vs. traffic load for the three difference schemes. For each color, high/medium/low opacity respectively denote the 50/95/99\%-tile delay.}
    \label{Fig:Contention_Delay}
    \end{subfigure}
    \begin{subfigure}[b]{0.99\columnwidth}
    \includegraphics[width=0.99\columnwidth]{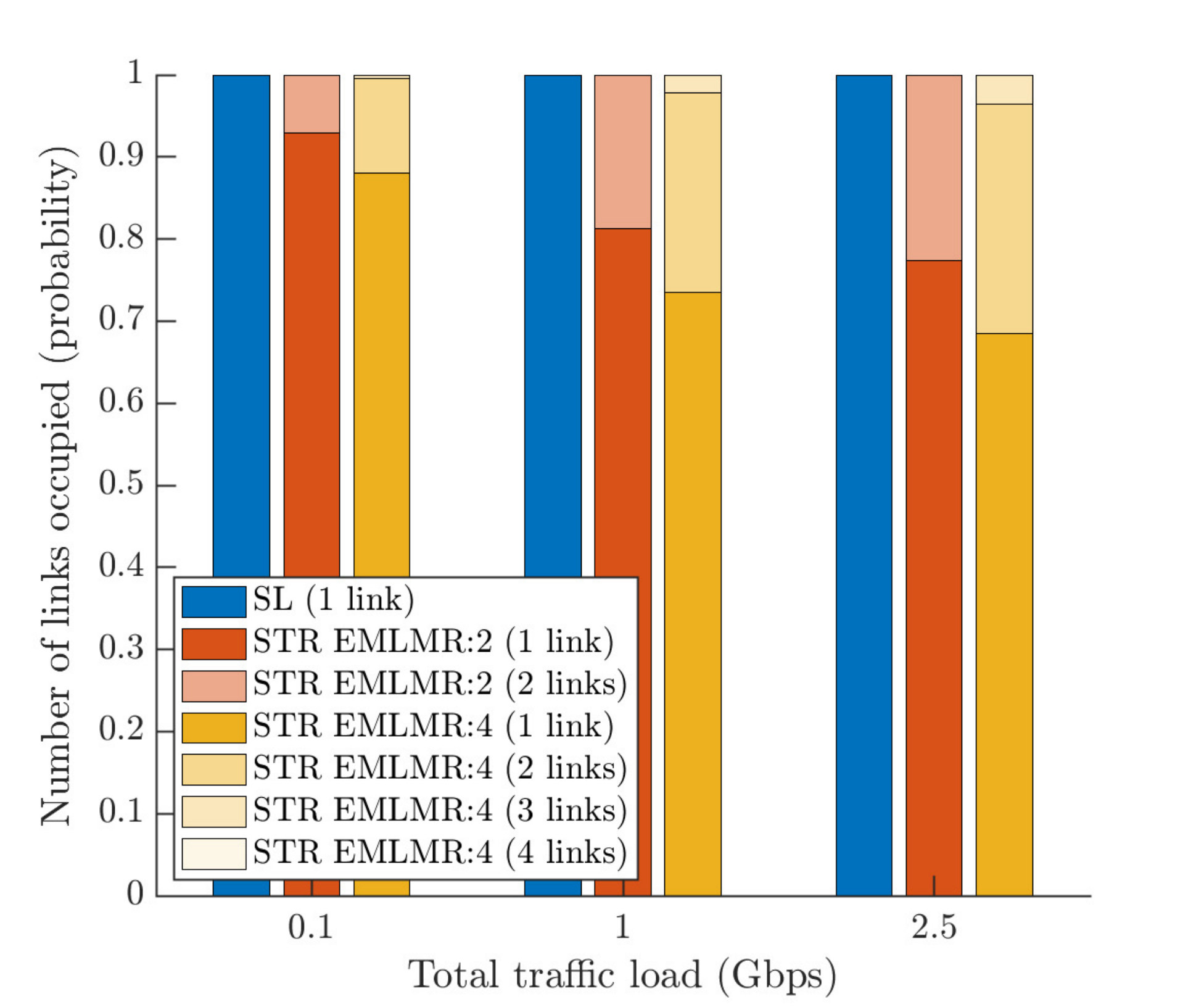}
    \caption{Probability for an active BSS to occupy a certain number of links concurrently vs. traffic load for the three different schemes.}
    \label{Fig:Contention_Occupancy}
    \end{subfigure}
    \caption{Crowded enterprise scenario: (a) delay and (b) number of links concurrently used by each active BSS vs. traffic load.}
    \label{Fig:Contention}
\end{figure}

In Fig.~\ref{Fig:Contention_Occupancy} we dig deeper into the delay anomaly by observing how STR EMLMR devices occupy the available links depending on their traffic load. The bars show, through different color opacity, the probability that an active BSS (i.e., with packets to transmit) will use a certain number of links concurrently. 
For a high traffic load of 2.5~Gbps, SL is limited to transmit on one link only, whereas STR EMLMR:2 employs a second interface 26\% of the time, and STR EMLMR:4 uses two or more interfaces 34\% of the time. A remarkable consequence (not shown for brevity) is that, with STR EMLMR:4, each contending BSS finds all four links occupied 24\% of the time. These events cause a deferral of the backoff countdown and prevent access to any wireless channel. 

In other words, while SL mode allows---or better said, forces---each BSS to operate on its own dedicated link 100\% of the time (Fig.~\ref{fs1}), whenever STR EMLMR BSSs use multiple links opportunistically they inevitably prevent at least another BSS from accessing at least one of its allocated channels (Figs.~\ref{fs2} and \ref{fs3}).

\subsubsection*{Packet aggregation}
Due to a higher contention, which results in longer backoff times, whenever a STR EMLMR device does succeed in accessing the channel, it must occasionally aggregate a larger number of queued packets as exemplified on the right hand side of Fig.~\ref{fig:diagram}. While not shown, we observed that for a load of 2.5~Gbps, switching from SL to STR EMLMR:4 decreases the median number of aggregated packets from 138 to 91, but it also increases its 99\%-tile value from 207 to 317. The latter corresponds to occasional intervals of long channel occupancy and undesirable delay anomalies.

\subsubsection*{Takeaway}
In the presence of high load and contention, STR EMLMR devices frequently access multiple links, thereby occasionally blocking contending neighbors for long periods of time and causing larger delays than those experienced by legacy SL under a static orthogonal channel allocation.

\section{Overcoming the Delay Anomaly}

How to side-step the delay anomaly occasionally 

experienced by STR EMLMR in crowded environments? In this section, we explore multiple practical options based on clever and/or extra channel assignment and compare their performance for the same enterprise scenario introduced in Section~IV: 
\begin{itemize}
    \item
    \textsl{EMLSR:2}, detailed in Section~II-A, with each MLD availing of two channels as in Fig.~\ref{fs2} but only equipped with one radio and thus only able to use one link at a time. This setup still requires a total of four channels.
    \item
    \textsl{STR EMLMR:1+1}, with each MLD using two links: one on a channel exclusively reserved (thus undergoing no contention)
    plus one on a channel shared with all other BSSs. This hybrid arrangement requires a total of five channels as opposed to the four required in Fig.~\ref{fs2}.
    \item
    \textsl{STR EMLMR:5}, with an overprovisioning of five links per MLD, each operating on a different channel, with all channels accessible by all four BSSs. Like the previous one, this setup requires a total of five channels, but it additionally requires five radio interfaces per MLD.
\end{itemize}

\begin{figure}
    \includegraphics[width=0.99\columnwidth]{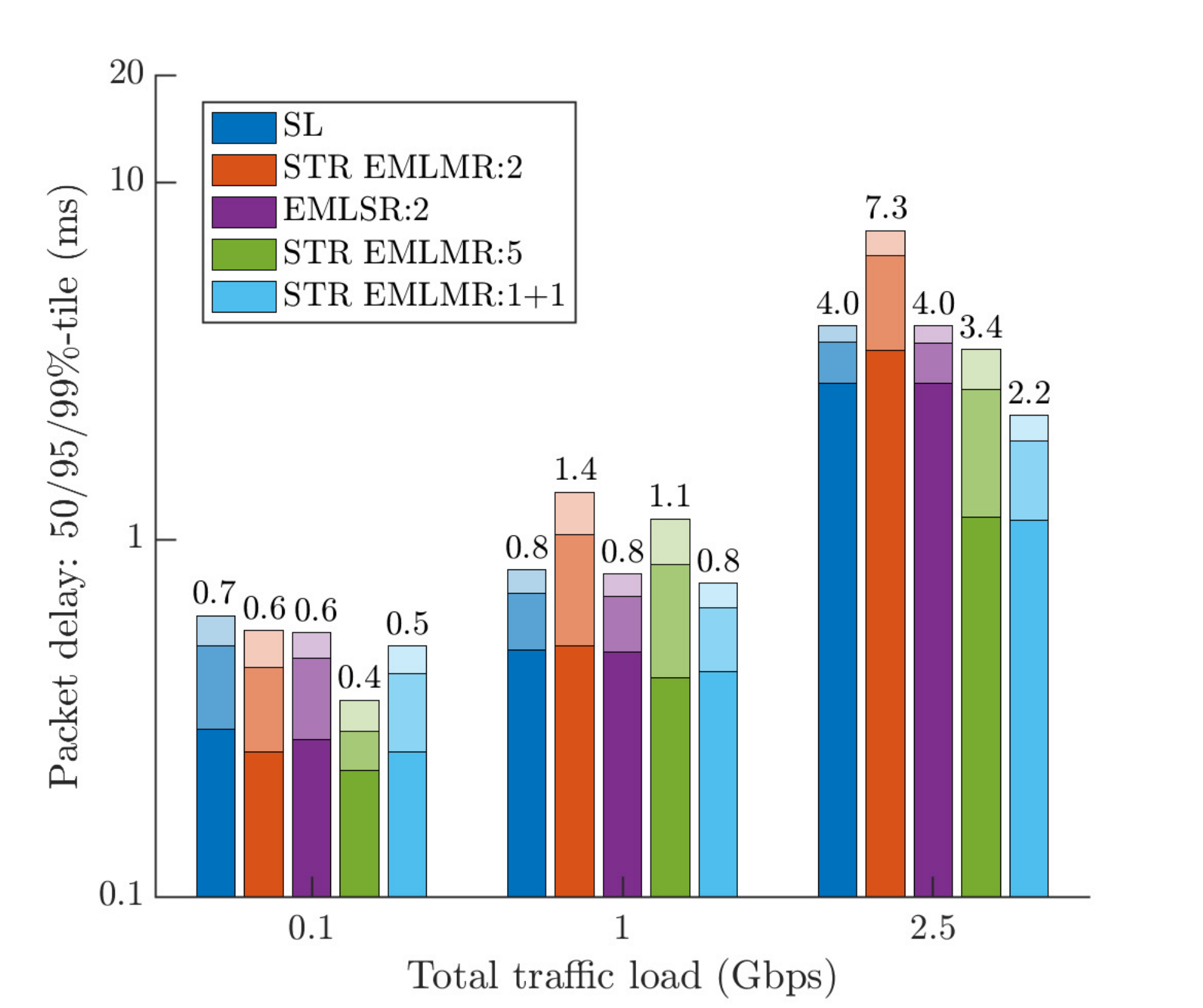}
    \caption{Crowded enterprise scenario: delay vs. traffic load for difference schemes. For each color, high/medium/low opacity respectively denotes the 50/95/99\%-tile delay.}
   
    \label{Fig:Solutions}
\end{figure}

Fig.~\ref{Fig:Solutions} displays the delay experienced by the three above approaches when compared to SL (Fig.~\ref{fs1}) and STR EMLMR:2 (Fig.~\ref{fs2}). We note how forcing each MLD to transmit on one link at a time with EMLSR:2 (purple) keeps the delay below or equal to that of SL (blue) across all values of load considered, while not increasing the total number of channels required.
At a load of 0.1~Gbps, all approaches experience low delays, with STR EMLMR:5 (green) achieving the lowest. Indeed, the low contention arising in this regime makes it likely for a MLD to encounter multiple links available, and juggling up to five running backoffs further reduces the delay. 
Interestingly, as the load grows to 2.5~Gbps, equipping each MLD with just two radios and operating STR EMLMR:1+1 (light blue) outperforms STR EMLMR:5, despite the latter employing as many as five radios per MLD. Indeed, delay reduction is owed not only to a larger number of channels and an increased system throughput, but also to circumventing the delay anomaly by guaranteeing one contention-free channel per BSS. 

Overall, to consistently outperform SL, STR EMLMR may thus require a total number of channels larger than the number of contending BSSs, and therefore equipping MLDs with additional radios, self-interference cancellation capabilities, and ensuring a sufficient inter-channel spacing. For instance, operating both in the 5~GHz and 6~GHz bands could allow accessing five 80~MHz channels with a spacing of 160~MHz.

\subsubsection*{Takeaway}
For consistent worst-case delay reduction, one may resort to STR EMLMR with more channels than contending BSSs and/or to performing a clever channel assignment that entirely circumvents delay anomalies caused by sporadic traffic starvation.

\section{Recap and Concluding Remarks}

Our study confirmed that in scenarios devoid of contention, STR EMLMR exploits extra links to transmit opportunistically, supporting significantly higher traffic loads (and therefore throughput) than SL while meeting strict delay requirements. Conversely, we discovered that in the presence of high load and contention, STR EMLMR devices frequently access multiple links, thereby blocking contending BSSs and occasionally causing larger delays than those experienced with a legacy SL operation with orthogonal channel assignment.

\subsubsection*{STR vs. NSTR EMLMR}
Though we focused on STR EMLMR for brevity, NSTR EMLMR too may incur delay anomalies. Indeed, its required alignment of simultaneous transmissions comes at the expense of spectrum reuse efficiency, ultimately creating higher contention and further increasing the chances that a certain BSS is prevented from accessing any channel.

\subsubsection*{Symmetric vs. asymmetric load}
While delay anomalies may arise under high traffic load across all contending APs, their likelihood and relevance are reduced when the traffic is unevenly distributed across contenders. Let us take the scenario studied in Section~III as an extreme example, with all traffic handled by a single active BSS. In such cases, STR EMLMR can efficiently map asymmetric traffic loads to all available links, drastically reducing the delay with respect to SL operations.

\subsubsection*{Single vs. multiple radios}
We observed that delay anomalies can be circumvented by employing EMLSR, which allows MLDs to opportunistically select a link among several but forces them to transmit on only one at a time. Although EMLSR makes for a lower complexity mode than EMLMR to reduce the delay at low traffic loads, using one link at a time prevents MLDs from achieving higher throughputs than under SL operation. While not shown in Fig.~\ref{Fig:NoContention}, a traffic load beyond 0.5~Gbps would 
eventually exceed the channel capacity of EMLSR, just as it does with SL. In this regime, availing of multiple radios would be the only approach to scale up the throughput so as to guarantee bounded delays.

\subsubsection*{Static vs. dynamic channel allocation}
We compared (i) a static channel assignment approach (SL, Fig.~\ref{fs1}), (ii) an entirely dynamic approach (STR EMLMR, Figs.~\ref{fs2} and \ref{fs3}), and (iii) a hybrid approach that cleverly reserves a certain channel for each BSS while leaving one more for contention (STR EMLMR:1+1). We found the latter to be most effective at reducing worst-case delays, even more so---and more economical---than equipping MLDs with more radio interfaces (STR EMLMR:5). Indeed, while STR EMLMR:1+1 guarantees at least one contention-free link for each BSS, STR EMLMR:5 merely spreads contention out over all available links, hence reducing the likelihood of a delay anomaly, but not necessarily overcoming it.

\section{Research Directions:\\Beyond Wi-Fi 7 and MLO}

The highly scenario-dependent performance of MLO, paired with the inherent uncertainty of the unlicensed spectrum, call for a learning-based optimization \cite{szott2022wi}. Suitably trained models could foresee the most appropriate MLO configuration for a set of overlapping BSSs, given input features such as the traffic load and its corresponding QoS requirements, preventing undesirable phenomena like the delay anomaly described in this paper.

Under the right implementation and design, and possibly evolving towards learning-based operations, MLO is bound to keep its promises of throughput augmentation and delay reduction by making the most efficient use of all available radio resources. 
Nonetheless, creating new transmission opportunities out of nowhere is not what MLO was meant for---a shortcoming that will eventually turn into a bottleneck as we, and our machines, demand more data delivered on time. With Wi-Fi 7 defined and MLO up and running, beyond-802.11be technologies are expected to conquer new frequency bands and/or boost the spatial reuse of the old ones through advanced multi-antenna AP coordination.

\subsubsection*{mmWave operations}
Expanding Wi-Fi operations into the 60 GHz band is an alluring prospect as it could increase the available spectrum by nearly one order of magnitude. And while 802.11 already defines modes of operation in this band (i.e., 802.11ad and 802.11ay), making mmWave Wi-Fi a greater commercial success may entail coming up with a more compatible design between sub- and above-8GHz operations, thereby reducing initial investment costs \cite{ResCor22}.

\subsubsection*{Multi-antenna AP coordination}
This paradigm is embodied by potential features like coordinated beamforming, aiming at packing APs with even more antennas not only to spatially multiplex their associated STAs, but also to suppress the interference generated/received to/from neighboring non-associated STAs. By exploiting multiple spatial degrees of freedom to place radiation nulls, coordinated beamforming could make neighboring BSSs invisible to each other, making inter-BSS contention a thing of the past and creating fertile ground for MLO to exploit \cite{garcia2021ieee}.

As we move towards Wi-Fi 8 with the formation of an 802.11 Task Group on Ultra High Reliability (UHR), above-8~GHz operations and AP coordination should not be added as a bolt-on, but rather conceived atop Wi-Fi 7 and embracing this MLO-native technology. On the one hand, a marriage between MLO and mmWave or AP coordination could be offering unprecedented and exciting challenges to our research community. But on the other, it creates the ultimate opportunity to meet the supreme goal of making unlicensed wireless the new wired. 
\typeout{}

	\bibliographystyle{IEEEtran}
\section*{Biographies}
\small

\noindent
\noindent
\textbf{Marc Carrascosa-Zamacois} is a Ph.D. candidate at Univ. Pompeu Fabra in Barcelona. His research interests are wireless networks, with a focus on performance optimization and latency minimization.

\vspace{0.2cm}
\noindent
\textbf{Giovanni Geraci} is an Assistant Professor at Univ. Pompeu Fabra in Barcelona and the Head of Telecommunications. He serves as an IEEE ComSoc Distinguished Lecturer, holds a dozen patents on wireless technologies, and received the IEEE ComSoc Outstanding Young Researcher Award for Europe, Middle East, and Africa.

\vspace{0.2cm}
\noindent
\textbf{Lorenzo Galati-Giordano} is a Senior Research Engineer at Nokia Bell Labs in Stuttgart, focusing on unlicensed spectrum technologies. He has 15+ years of academic and industrial experience and has co-authored tens of commercial patents, publications, and IEEE 802.11 standard contributions.

\vspace{0.2cm}
\noindent
\textbf{Anders Jonsson} is a Full Professor at Univ. Pompeu Fabra in Barcelona and the Head of the Artificial Intelligence and Machine Learning research group. His research interests involve sequential decision problems, automated planning, and reinforcement learning.

\vspace{0.2cm}
\noindent
\textbf{Boris Bellalta} is a Full Professor at Univ. Pompeu Fabra in Barcelona and the Head of the Wireless Networking research group. His research interests are in wireless networks, adaptive systems, machine learning and eXtended Reality.

\end{document}